# Movable Antenna Assisted OAM Wireless Communications With Misaligned Transceiver

Hongyun Jin, Wenchi Cheng, Haiyue Jing, and Jingqing Wang

State Key Laboratory of Integrated Services Networks, Xidian University, Xi'an, China

E-mail: {*hongyunjin@stu.xidian.edu.cn, wccheng@xidian.edu.cn, hyjing@stu.xidian.edu.cn, jqwangxd@xidian.edu.cn*}

*Abstract*—The vortex electromagnetic wave carried by multiple orthogonal orbital angular momentum (OAM) modes in the same frequency band can be applied to the field of wireless communications, which greatly increases the spectrum efficiency. The uniform circular array (UCA) is the classical structure to generate and receive vortex electromagnetic waves with multiple OAM-modes. However, when the transmit and receive UCAs are misaligned, there will be interference among the OAM-modes and the signal cannot be recovered at the receiver. In order to solve this problem, we propose movable antenna (MA) assisted OAM wireless communications scheme. We estimate the rotation angle between transmit and receive UCAs and feed it back to the transmitter. Then, the MA at the transmitter adjusts the rotation angle to achieve alignment of the UCA at both the receiver and transmitter. Simulation results show that our scheme can significantly improve the spectrum efficiency.

## I. INTRODUCTION

IN the field of wireless communication, information is carried through various dimensions of electromagnetic waves, including frequency, amplitude, phase, and polarization. Orbital angular momentum (OAM) as a promising technology for next-generation wireless communication, characterized by properties distinct from the linear momentum of electromagnetic wave radiation [1]. It is expected to be utilized for expanding communication dimensions, where the new dimension can be employed for data transmission or as a new degree of freedom for beamforming control, thereby increasing transmission capacity and enhancing system performance [2]–[4].

The uniform circular array (UCA), as the classical structure to generate and receive vortex waves with multiple OAM-modes, offers notable advantages in terms of flexibility and convenience in generating and receiving multiple OAM-modes [5], [6]. The authors of [7] verified that among radial array, tangential array, and UCA, UCA is the best way to generate OAM beams from the perspective of generation and demultiplexing. The authors of [8] verified that UCA can generate OAM-modes OAM radio beam by using RF analog synthesis method and baseband digital synthesis method. The author of [9] analyzed the theory of generating OAM by UCA and designed a UCA that can generate pure different OAM-modes. The authors of [10] showed the UCA schematic configuration for generating OAM beams and analyzed the impact of array error on the radiation field.

Movable antenna (MA) technology is a recent development that fully exploits the wireless channel spatial variation in a confined region by enabling local movement of the antenna. Specifically, the positions of antennas at the transmitter and/or receiver can be dynamically changed to obtain better channel conditions for improving the communication performance [11]. With the capability of flexible movement, the MAs can be deployed at positions with more favorable channel conditions in the spatial region to achieve higher spatial diversity gains. The authors of [12] proposed a new multiple-input multiple-output (MIMO) communication system with MAs to exploit the antenna position optimization for enhancing the capacity. The authors of [13] focus to maximize the effective received signal-to-noise ratio by jointly optimizing the transmit beamforming and the positions of the MA array.

However, it is unrealistic to ensure the alignment of the transmit and receive UCAs in the application of the actual wireless communication system [14]. Due to the misalignment of the UCAs at the transmitter and receiver, the phase turbulence will result in significant interference among OAM-modes, and the signal cannot be recovered at the receiver [15]. The author of [16] proposed a mode decomposition method for the scene with the UCA transceivers are non-coaxial. The author of [17] proposed a channel independent beamforming and the corresponding fast symbol-wise maximum-likelihood detection, where the transmit and the receive UCAs are non-coaxial with each other. The authors of [18] aims at the problem that different OAM-modes cannot be solved in the case of antenna misalignment, proposed a joint beamforming and pre-detection scheme to turn the channel matrix into a circular matrix.

In order to solve the above problems, we propose the movable antenna assisted OAM wireless communications with misaligned transceiver. According to the system model, we derived the expression of channel gain, and solved the relationship between the rotation angle and the channel gain. We adjust the rotation angle between UCAs by feeding back the calculated antenna rotation angle to the transmitter to align the UCAs. We estimate the channel matrix by minimum mean square error (MMSE), then decompose the channel matrix and obtain the expression of rotation angle through joint calculation. For MA-assisted OAM wireless communications, we can jointly optimize the relative positions of the transmit and receive UCAs to maximize channel capacity. To achieve high channel capacity in OAM wireless communications, multiple data streams are typically transmitted simultaneously. For misaligned OAM wireless communications, it is necessary to optimize the relative positions of the transmite and receive UCAs to eliminate interference between OAM-modes. According to the simulation results, our method can significantly improve the spectrum efficiency (SE).

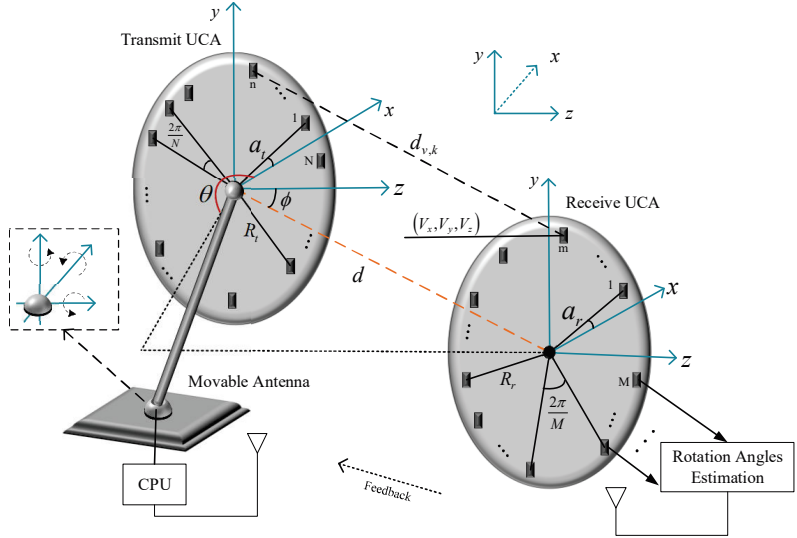

Fig. 1. The system model of movable antenna assisted OAM wireless communications with misaligned transmit and receive UCAs.

The rest of this paper is organized as follows. Section II gives the system model of movable antenna assisted OAM wireless communications with misaligned transmit and receive UCAs. Section III estimates the channel matrix, derive the expression of channel gain, and obtain the expression of deflection angle through joint solution. Section IV provides the numerical results. The paper concludes with Section V.

Notation: Matrices and vectors are denoted by the capital letters and the lowercase letters in bold, respectively.

## II. THE SYSTEM MODEL OF MOVABLE ANTENNA ASSISTED OAM WIRELESS COMMUNICATIONS WITH MISALIGNED TRANSMIT AND RECEIVE UCAS

Figure 1 depicts the system model of movable antenna assisted OAM wireless communications with misaligned transmit and receive UCAs. The distance between the centers of the transmit and receive UCAs is indicated by $d$. We denote by $d_{vk}$ the distance from the $k$th ($0 \leq k \leq K-1$) array-element on transmit UCA to the $v$th ($0 \leq v \leq V-1$) array-element on receive UCA. The array-elements on the UCAs are uniformly arranged around the circumference. There are $K$ array-elements on the transmit UCA and $V$ array-elements on the receive UCA, respectively. The azimuth angles of two adjacent array-elements on the transmit UCA and receive UCA are $2\pi/K$ and $2\pi/V$, respectively. We denote by $a_t$ and $a_r$ the angle between the azimuth of the first array-element and the zero radian corresponding to the transmit and receive UCAs, respectively. We denote by $\varphi_k = 2\pi(k-1)/K$ the basic azimuth for the transmit UCA, $\varphi_k + a_t$ the phase angle of the $k$th array-element on the transmit UCA. We denote by $\psi_v = 2\pi(v-1)/V$ the basic azimuth for the receive UCA, $\psi_v + a_r$ the phase angle of the $v$th array-element on the receive UCA. For the transmit UCA, each array-element is loaded with the same signal but with a continuous phase delay between adjacent array-elements, and the phase of the signal loaded from the first array-element to the last array-element increases in a fixed gradient to $2\pi l$, where $l$ represents the OAM-modes. The radius of the transmit UCA and receive UCA are $R_t$ and $R_r$, respectively. We establish the spatial right-angle coordinate system as shown in the Figure 1, where the coordinate origin is the center of the transmit UCA.

The parameter $\theta_t$ represents the angle between the $x$-axis and the projection of the line from the center of the transmit UCA to the center of the receive UCA. The parameter $\phi_t$ denotes the angle between the $z$-axis and the line from the center of the transmit UCA to the center of the receive UCA. After the signals are received at the receiver UCA, we estimate the channel information and then perform the calculation of the angle parameters. Then, the angle parameters are feed back to the MA at the transmitter. The MA makes the corresponding angle adjustment to ensure that the transmitter and receiver UCAs are aligned in parallel. The MA is controlled by the central processing unit (CPU), which can be independently adjusted in terms of three-dimensional rotation. Different from the traditional OAM wireless communications with fixed-position antennas, our proposed system can flexibly change the positions of the transmite UCA, thereby reshaping the channel matrix between UCAs to improve the spectrum efficiency.

## III. Joint channel and angle estimation for unaligned vortex electromagnetic wave wireless communications

The modulated signal located at the $k$th array-element on the transmit UCA, denoted by $x_k$, can be given as follows:

$$x_k = \sum_{l=0}^{K-1} \frac{s_l}{\sqrt{K}} e^{j(\frac{2\pi k}{K}+a_t)l}, \quad (1)$$

where $s_l$ denotes the transmit signal corresponding to OAM-mode $l$, $K$ is the number of available OAM-modes, and $e^{j(\frac{2\pi k}{K}+a_t)l}$ are the phase shifts of UCA modulations.

The channel gain between the $k$th array-element on the transmit UCA and the $v$th array-element on the receive UCA is denoted by $h_{v,k}$ and can be given as follows:

$$h_{v,k} = \frac{\beta \lambda e^{-j\frac{2\pi}{\lambda} d_{v,k}}}{4\pi d_{v,k}}, \quad (2)$$

where $\beta$ denotes the parameter gathering relevant constants on antenna array-elements, such as the reduction and phase rotation brought by the antenna at both ends, $d_{v,k}$ is the transmission distance from the $k$th array-element on the transmit UCA to the $v$th array-element on the receive UCA, $\lambda$ represents the wavelength of the signal.

The coordinate of the $k$th array-element on the transmit UCA, denoted by $(K_x, K_y, K_z)$, can be given as follows:

$$\begin{cases} K_x = R_t \cos(\varphi_k + a_t); \\ K_y = R_t \sin(\varphi_k + a_t); \\ K_z = 0. \end{cases} \quad (3)$$

The coordinate of the center of the receive UCA is $(d\sin\phi_t \cos\theta_t, d\sin\phi_t \sin\theta_t, d\cos\phi_t)$. Then, the coordinate of the $v$th array-element on the receive UCA, denoted by $(V_x, V_y, V_z)$, can be given as follows:

$$\begin{cases} V_x = d\sin\phi_t \cos\theta_t + R_r \cos(\psi_v + a_r); \\ V_y = d\sin\phi_t \sin\theta_t + R_r \sin(\psi_v + a_r); \\ V_z = d\cos\phi_t. \end{cases} \quad (4)$$

Then, the notation $d_{v,k}$ can be given as follows:

$$d_{v,k} = \sqrt{[V_x - K_x]^2 + [V_y - K_y]^2 + V_z^2} \\ = \sqrt{d^2 + R_r^2 + R_t^2 + 2\mathscr{A}_v + 2\mathscr{B}_{vk} + 2\mathscr{C}_k}, \quad (5)$$

where $\mathscr{A}_v$, $\mathscr{B}_{vk}$, and $\mathscr{C}_k$ are given as follows:

$$\begin{cases} \mathscr{A}_v = dR_t \sin\phi_t \cos(\psi_v + a_r - \theta_t); \\ \mathscr{B}_{vk} = -R_r \cos(\psi_v - \varphi_k + a_t - a_r); \\ \mathscr{C}_k = -dR_t \sin\phi_t \cos(\varphi_k + a_r - \theta_t). \end{cases} \quad (6)$$

To make an approximate substitution, we rewrite $d_{v,k}$ as follows:

$$d_{v,k} = \sqrt{d^2 + R_r^2 + R_t^2} \left( \sqrt{1 + \frac{2(\mathscr{A}_v + \mathscr{B}_{vk} + \mathscr{C}_k)}{d^2 + R_r^2 + R_t^2}} \right). \quad (7)$$

Since $d >> R_r$ and $d >> R_t$, we can approximate $d_{v,k}$ as $\tilde{d}_{v,k}$ using $\sqrt{1+2x} \approx 1+x$, which can be given as follows:

$$\tilde{d}_{vk} = \sqrt{d^2 + R_r^2 + R_t^2} + \frac{\mathscr{A}_v + \mathscr{B}_{vk} + \mathscr{C}_k}{\sqrt{d^2 + R_r^2 + R_t^2}}. \quad (8)$$

Then, the notation $h_{v,k}$ can be approximated as $\tilde{h}_{v,k}$, which can be given as follows:

$$\tilde{h}_{v,k} = \frac{\beta \lambda e^{-j\frac{2\pi}{\lambda}\tilde{d}_{v,k}}}{4\pi d} \\ = \frac{\beta \lambda}{4\pi d} e^{-j\frac{2\pi}{\lambda}(\sqrt{d^2+R_r^2+R_t^2} + \frac{\mathscr{A}_v + \mathscr{B}_{vk} + \mathscr{C}_k}{\sqrt{d^2+R_r^2+R_t^2}})} \\ = \frac{\beta \lambda}{4\pi d} e^{-j\frac{2\pi}{\lambda}\sqrt{d^2+R_r^2+R_t^2}} (e^{\mathscr{A}_v} e^{\mathscr{B}_{vk}} e^{\mathscr{C}_k})^{\frac{-j2\pi}{\lambda\sqrt{d^2+R_r^2+R_t^2}}}. \quad (9)$$

For the convenience of subsequent calculations, we denote by $\mathscr{D}$ and $\mathscr{F}$ as follows:

$$\begin{cases} \mathscr{D} = \frac{\beta \lambda}{4\pi d} e^{-j\frac{2\pi}{\lambda}\sqrt{d^2+R_r^2+R_t^2}}; \\ \mathscr{F} = \frac{-j2\pi}{\lambda\sqrt{d^2+R_r^2+R_t^2}}. \end{cases} \quad (10)$$

Then, we further write $\tilde{h}_{v,k}$ as follows:

$$\tilde{h}_{v,k} = \mathscr{D}(e^{\mathscr{A}_v})^{\mathscr{F}} (e^{\mathscr{B}_{vk}})^{\mathscr{F}} (e^{\mathscr{C}_k})^{\mathscr{F}}. \quad (11)$$

We obtain the channel estimation matrix $\hat{\boldsymbol{H}}$ using the MMSE shown as follows:

$$\hat{\boldsymbol{H}} = \begin{bmatrix} \hat{h}_{0,0} & \cdots & \hat{h}_{0,k} & \cdots & \hat{h}_{0,K-1} \\ \vdots & \ddots & \vdots & \ddots & \vdots \\ \hat{h}_{v,0} & \cdots & \hat{h}_{v,k} & \cdots & \hat{h}_{v,K-1} \\ \vdots & \ddots & \vdots & \ddots & \vdots \\ \hat{h}_{V-1,0} & \cdots & \hat{h}_{V-1,k} & \cdots & \hat{h}_{V-1,K-1} \end{bmatrix}. \quad (12)$$

Substituting the $\hat{h}_{v,k}$ of the $v$th row and $k$th column of the channel estimation matrix $\hat{\boldsymbol{H}}$ into Eq. (11), we further obtain:

$$\frac{1}{\mathscr{D}} (e^{\mathscr{A}_v})^{-\mathscr{F}} \hat{h}_{v,k} (e^{\mathscr{C}_k})^{-\mathscr{F}} = (e^{\mathscr{B}_{vk}})^{\mathscr{F}}. \quad (13)$$

For transmit and receive UCAs, the array-elements are arranged uniformly around the circumference. So, we can find the values of the array-element azimuthal intervals of $\frac{\pi}{2}$, $\pi$, and $\frac{3\pi}{2}$ in Eq. (13). According to the periodicity of trigonometric functions, we can find multiple pairs of corresponding values, substituting them into Eq. (13), solving them by association, and then simplifying them to obtain the rotation angles.

We substitute $\hat{h}_{v,\frac{K}{2}+1}$, $\hat{h}_{v+1,\frac{3K}{4}+1}$ into Eq. (13) and obtain:

$$\begin{cases} \frac{1}{\mathscr{D}} \hat{h}_{v,\frac{K}{2}+1} e^{\cos(\theta-\alpha)\sin\phi \mathscr{F} d(R_t+R_r)} = e^{R_t R_r \mathscr{F}}; \\ \frac{1}{\mathscr{D}} \hat{h}_{v+1,\frac{3K}{4}+1} e^{-\sin(\theta-\alpha)\sin\phi \mathscr{F} d(R_t+R_r)} = e^{R_t R_r \mathscr{F}}. \end{cases} \quad (14)$$

*Proof:* See Appendix A. ∎

We consider the angle between the phase angle of the first array-element on the UCAs at both ends and the corresponding zero radians to be equal, i.e., $a = a_r = a_t$. By solving Eq. (14) jointly, we can obtain the following angle parameter:

$$\tan(\theta - \alpha) = \frac{\ln \frac{\mathscr{D}}{\hat{h}_{v+1,\frac{3K}{4}+1}} + R_t R_r \mathscr{F}}{\ln \frac{\mathscr{D}}{\hat{h}_{v,\frac{K}{2}+1}} + R_t R_r \mathscr{F}}. \quad (15)$$

Then, we denote by $\delta_\theta = \tan(\theta - \alpha)$ and obtain the angle parameter $\sin\phi$ as follows:

$$\sin\phi = \frac{\ln\frac{\hat{h}_{v,\frac{K}{2}+1}}{\mathcal{D}} - R_t R_r \mathcal{F}}{\cos(\arctan\delta_\theta)\mathcal{F}(R_t + R_r)d}. \quad (16)$$

Based on Eqs. (10), (15), and (16), we can obtain the rotation angles in Eq. (17).

The signal received by the $v$th array-element on the receive UCA is represented by $r_v$ as follows:

$$\begin{aligned} r_v &= \sum_{k=0}^{K-1} \tilde{h}_{v,k} x_k + z_v \\ &= \sum_{k=0}^{K-1} \tilde{h}_{v,k} \sum_{l=0}^{K-1} \frac{1}{\sqrt{K}} s_l e^{j(\varphi_k + a_t)l} + z_v \\ &= \sum_{l=0}^{K-1} s_l \tilde{h}_{v,l} + z_v, \end{aligned} \quad (18)$$

where $z_v$ represents the noise of the $v$th array-element on the receive UCA. We denote by $\tilde{h}_{v,l}$ the channel gain from the transmit UCA to the $v$th array-element on the receive UCA corresponding to the $l$th OAM-mode, which can be given as follows:

$$\tilde{h}_{v,l} = \sum_{k=0}^{K-1} \mathcal{D}(e^{\mathcal{A}_v})^{\mathcal{F}} (e^{\mathcal{B}_{vk}})^{\mathcal{F}} (e^{\mathcal{C}_k})^{\mathcal{F}} e^{j(\varphi_k + a_t)l}. \quad (19)$$

The SE of our OAM wireless communications can be derived as follows:

$$SE = \sum_{l=0}^{K-1} \log_2\left(1 + \frac{\sum_{v=0}^{V-1} |\tilde{h}_{v,l}|^2 |s_l|^2}{\sigma_l^2}\right), \quad (20)$$

where $\sigma_l^2$ denotes the noise variance corresponding to the $l$th OAM-mode.

## IV. PERFORMANCE EVALUATIONS

In this section, numerical simulation results are presented to evaluate the performance of our developed MA assisted OAM wireless communication with non-aligned transceiver. We compare the spectrum efficiency corresponding to MA assisted OAM wireless communications with those of OAM wireless communications without MA assisted. Then, we evaluate the spectrum efficiencies for different different numbers of array-elements versus the included angle $\theta$ and $\phi$, respectively. The propagation environment and the positions of antennas determine the channel response for MA systems. In this paper, we assume that the size of the region for moving the antenna is much smaller than the propagation distance between the transmit UCA and receive UCA, so that the far-field condition is satisfied. Throughout the evaluation, we set the communication system operating at the 5.8 GHz frequency band and the distance as $20\lambda$. We also set $R = r = 0.5m$, $\beta = 1$, and $a_r = a_t = 0$.

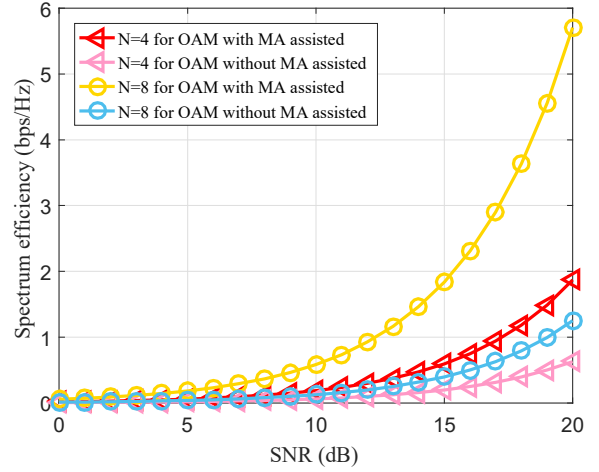

Fig. 2. The spectrum efficiencies for MA-assisted OAM wireless communications and OAM wireless communications without MA assistance.

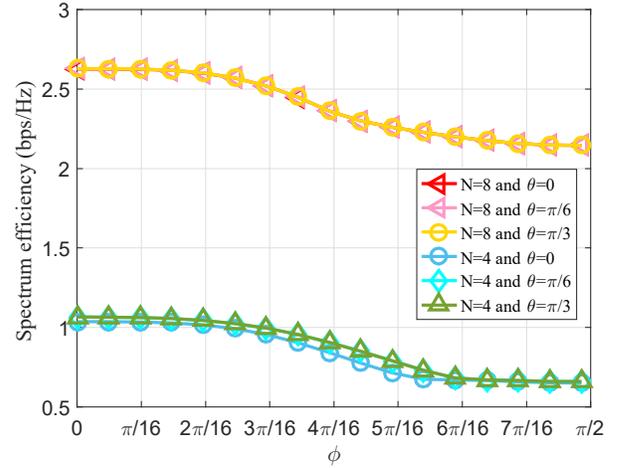

Fig. 3. The obtained spectrum efficiency versus the included angles $\phi$ and $\theta$.

Figure 2 depicts the spectrum efficiencies for MA-assisted OAM wireless communications and OAM wireless communications without MA assistance, where the transmit and receive UCA are non-aligned. We can obtain that the spectrum efficiency of MA-assisted OAM wireless communications is much larger than that of OAM wireless communications without MA assistance. This is because the MA-assisted OAM wireless communication can adjust misaligned antennas to alignment, thereby eliminating interference among OAM-modes. Using the MA-assisted OAM wireless communications, the signals carried by different OAM-modes are interference-free. The spectrum efficiencies of MA-assisted OAM wireless communications increase as the number of array-elements increases.

Figure 3 illustrates the spectrum efficiency of MA-assisted OAM wireless communication versus the included angle $\phi$ and $\theta$. With the variation of $\phi$ and $\theta$, the change of spectrum efficiency is not significant, which is attributed to the effec-

$$\begin{cases} \theta = \arctan\left(\dfrac{\lambda\sqrt{d^2+R_r{}^2+R_t{}^2}\ln\dfrac{\beta\lambda}{4\pi d\hat{h}_{v+1,\frac{3K}{4}+1}}-2\pi j(R_r{}^2+R_t{}^2+d^2+R_rR_t)}{\lambda\sqrt{d^2+R_r{}^2+R_t{}^2}\ln\dfrac{\beta\lambda}{4\pi d\hat{h}_{v,\frac{K}{2}+1}}-2\pi j(R_r{}^2+R_t{}^2+d^2+R_rR_t)}\right)+a; \\ \phi = \arcsin\left(\dfrac{\dfrac{j\lambda\sqrt{d^2+R_r{}^2+R_t{}^2}}{2\pi}\ln\dfrac{4\pi d\hat{h}_{v,\frac{K}{2}+1}}{\beta\lambda}-(R_r{}^2+R_t{}^2+d^2+R_rR_t)}{\cos(\arctan\delta_\theta)d(R_t+R_r)}\right). \end{cases} \quad (17)$$

tive adjustment achieved by the MA-assisted OAM wireless communication, thereby maintaining high spectrum efficiency. For MA-assisted OAM wireless communication with a larger number of array-elements, MA exhibits better adjustment performance. When the included angles $\theta$ and $\phi$ are equal to 0, i.e., the transmit and receive UCAs are aligned with other, we have the maximum spectrum efficiency.

## V. Conclusions

In this paper, we proposed the movable antenna assisted OAM wireless communications where the transmit and receive UCAs are misaligned. We estimate the rotation angle between transmit and receive UCAs and feed it back to the transmitter. Then, the MA at the transmitter adjusts the rotation angle to achieve alignment of the UCA at both the receiver and transmitter. The adoption of MA provides a novel solution to align the transceiver antennas for achieving more practical OAM wireless communications with misaligned transceiver. Simulation results validated that the spectrum efficiency of movable antenna assisted OAM wireless communications is much larger than that of OAM wireless communications without movable antenna assisted.

## Appendix A

$$(e^{\mathscr{A}_v})^{\mathscr{F}} = \begin{bmatrix} e^{\cos(\theta-a_r)} & & & \\ & \ddots & & \\ & & e^{\sin(\theta-a_r)} & \\ & & & \ddots \\ & & & & e^{\cos(\theta-a_r+\frac{2\pi}{V})} \end{bmatrix}^{dR_r\mathscr{F}\sin\phi} \quad (21)$$

$$(e^{\mathscr{C}_k})^{\mathscr{F}} = \begin{bmatrix} e^{\cos(\theta-a_t)} & & & \\ & \ddots & & \\ & & e^{\sin(\theta-a_t)} & \\ & & & \ddots \\ & & & & e^{\cos(\theta-a_t+\frac{2\pi}{K})} \end{bmatrix}^{-dR_t\mathscr{F}\sin\phi} \quad (22)$$

## References

[1] W. Cheng, W. Zhang, H. Jing, S. Gao, and H. Zhang, "Orbital angular momentum for wireless communications," *IEEE Wireless Communications*, vol. 26, no. 1, pp. 100–107, Feb. 2019.

[2] Y. Wang, X. Sun, and L. Liu, "Millimeter-wave orbital angular momentum: Generation, detection, and applications: A review on millimeter wave orbital angular momentum antennas," *IEEE Microwave Magazine*, vol. 25, no. 1, pp. 37–57, 2024.

[3] H. Jin, W. Cheng, and H. Jing, "Quasi-fractal UCA based two-dimensional OAM orthogonal transmission," in *GLOBECOM 2023 - 2023 IEEE Global Communications Conference*, 2023, pp. 7611–7616.

[4] M. Zhu, J. Zhang, B. Hua, M. Lei, Y. Cai, L. Tian, D. Wang, W. Xu, C. Zhang, Y. Huang, J. Yu, and X. You, "Ultra-wideband fiber-THz-fiber seamless integration communication system toward 6G: architecture, key techniques, and testbed implementation," *Science China (Information Sciences)*, vol. 66, no. 01, pp. 296–313, 2023.

[5] T. Yuan, Y. Cheng, H. Wang, and Y. Qin, "Mode characteristics of vortical radio wave generated by circular phased array: Theoretical and experimental results," *IEEE Transactions on Antennas and Propagation*, vol. 65, no. 2, pp. 688–695, 2017.

[6] W. Yu, B. Zhou, Z. Bu, and S. Wang, "Analyze UCA based OAM communication from spatial correlation," *IEEE Access*, vol. 8, pp. 194 590–194 600, 2020.

[7] H. Wu, Y. Yuan, Z. Zhang, and J. Cang, "UCA-based orbital angular momentum radio beam generation and reception under different array configurations," in *Sixth International Conference on Wireless Communications and Signal Processing (WCSP)*, Oct. 2014, pp. 1–6.

[8] R. Chen, W.-X. Long, X. Wang, and L. Jiandong, "Multi-mode oam radio waves: Generation, angle of arrival estimation and reception with UCAs," *IEEE Transactions on Wireless Communications*, vol. 19, no. 10, pp. 6932–6947, 2020.

[9] M. Lin, Y. Gao, P. Liu, and J. Liu, "Theoretical analyses and design of circular array to generate orbital angular momentum," *IEEE Transactions on Antennas and Propagation*, vol. 65, no. 7, pp. 3510–3519, 2017.

[10] K. Liu, H. Liu, Y. Qin, Y. Cheng, S. Wang, X. Li, and H. Wang, "Generation of OAM beams using phased array in the microwave band," *IEEE Transactions on Antennas and Propagation*, vol. 64, no. 9, pp. 3850–3857, Sep. 2016.

[11] L. Zhu, W. Ma, and R. Zhang, "Movable antennas for wireless communication: Opportunities and challenges," *IEEE Communications Magazine*, pp. 1–7, 2023.

[12] W. Ma, L. Zhu, and R. Zhang, "MIMO capacity characterization for movable antenna systems," *IEEE Transactions on Wireless Communications*, vol. 23, no. 4, pp. 3392–3407, 2024.

[13] G. Hu, Q. Wu, J. Ouyang, K. Xu, Y. Cai, and N. Al-Dhahir, "Movable-antenna-array-enabled communications with CoMP reception," *IEEE Communications Letters*, vol. 28, no. 4, pp. 947–951, 2024.

[14] R. Chen, H. Xu, M. Moretti, and J. Li, "Beam steering for the misalignment in UCA-based OAM communication systems," *IEEE Wireless Communications Letters*, vol. 7, no. 4, pp. 582–585, Aug. 2018.

[15] W. Cheng, H. Zhang, L. Liang, H. Jing, and Z. Li, "Orbital-Angular-Momentum embedded massive MIMO: Achieving multiplicative spectrum-efficiency for mmwave communications," *IEEE Access*, vol. 6, pp. 2732–2745, 2018.

[16] H. Jing, W. Cheng, X.-G. Xia, and H. Zhang, "Orbital-angular-momentum versus MIMO: Orthogonality, degree of freedom, and capacity," in *2018 IEEE 29th Annual International Symposium on Personal, Indoor, and Mobile Radio Communications (PIMRC)*, Sep. 2018, pp. 1–7.

[17] H. Jing, W. Cheng, and X.-G. Xia, "A simple channel independent beamforming scheme with parallel uniform circular array," *IEEE Communications Letters*, vol. 23, no. 3, pp. 414–417, 2019.

[18] W. Cheng, H. Jing, W. Zhang, Z. Li, and H. Zhang, "Achieving practical OAM based wireless communications with misaligned transceiver," in *ICC 2019 - 2019 IEEE International Conference on Communications (ICC)*, 2019, pp. 1–6.